\documentclass[prd,nofootinbib,tightenlines,showkeys,showpacs]{revtex4}
\usepackage{amsfonts}
\usepackage{amssymb}
\usepackage{amsmath}
\usepackage{indentfirst}
\usepackage[latin1]{inputenc}

\begin{document}

\newcommand{\be}{\begin{equation}} \newcommand{\ee}{\end{equation}}
\newcommand{\bea}{\begin{eqnarray}}\newcommand{\eea}{\end{eqnarray}}

\title{{\small\hfill IMSc/2009/02/04}\\
{\small\hfill SINP/TNP/2009/10 }\\
Deformed Oscillator Algebras and QFT in $\kappa$-Minkowski Spacetime}

\author{T. R. Govindarajan \footnote{trg@imsc.res.in}}

\affiliation{Institute of Mathematical Sciences, CIT Campus, Taramani, Chennai 600113, India}

\author{Kumar S. Gupta \footnote {kumars.gupta@saha.ac.in}}

\affiliation{Theory Division, Saha Institute of Nuclear Physics, 1/AF
Bidhannagar, Calcutta 700064, India}

\author{E. Harikumar \footnote{harisp@uohyd.ernet.in}}

\affiliation{School of Physics, University of Hyderabad, Hyderabad 500046, India}

\author{S. Meljanac \footnote {meljanac@irb.hr} and D. Meljanac
\footnote{dmeljan@irb.hr}}

\affiliation{Rudjer Bo\v{s}kovi\'c Institute, Bijeni\v cka  c.54, HR-10002
Zagreb, Croatia}

\vspace*{1cm}

\begin{abstract}
In this paper we study the deformed statistics and oscillator algebras of quantum fields defined in $\kappa$-Minkowski spacetime. The twisted flip operator obtained from the twist associated with the star product requires an enlargement of the Poincar\'e algebra to include the dilatation generators. Here we propose a novel notion of a fully covariant flip operator and show that to the first order in the deformation parameter it can be expressed completely in terms of the Poincar\'e generators alone. The $R$-matrices corresponding to the twisted and the covariant flip operators are compared up to first order in the deformation parameter and they are shown to be different. We also construct the deformed algebra of the creation and annihilation operators that arise in the mode expansion of a scalar field in $\kappa$-Minkowski spacetime. We obtain a large class of such new deformed algebras which, for certain choice of realizations, reduce to results known in the literature. 

\end{abstract} 

\pacs{ 11.10.Nx, 11.30.Cp}

\keywords {$\kappa$ deformed space, noncommutative geometry, statistics }

\maketitle

\section{ Introduction}
Noncommutative geometry as well as formulation and study of physical theories on noncommutative spaces have been attracting wide attention for quite some time now \cite{rev,rev1,rev2,rev3,rev4,st1,st2,st3,st4,st5}. Unique features of the theories on such spaces and also the fact that noncommutative geometry provides one of the possible approaches for describing Planck scale physics, notably, quantum gravity are some of the motivations for the renewed interest in these studies \cite{dop, connes, wess}. A simple and by now reasonably well studied model of noncommutative space is the Moyal space. One of the interesting aspects brought out by these studies is the role of Hopf algebra (quantum group) \cite{majid} in analyzing the symmetries of field theories on Moyal space. Though, in the conventional sense, the Lorentz symmetry is lost in these theories, it is now well understood that using Hopf algebra approach, Lorentz invariance can be retained in these noncommutative models, enabling the conventional interpretation of field quanta \cite{chaichan}.

In the Hopf algebra approach, the underlying symmetry algebra of the noncommutative theory acts on multi-particle states through the twisted coproducts of the symmetry generators. Alternate attempts to construct gravity theories on noncommutative spaces, where a compatibility between the so called $*$-product and the action of diffeomorphism symmetry generators also led to the introduction of twisted Leibniz rule (i.e. coproducts) for these generators \cite{wess}.

Noncommutative spaces which are more general than Moyal space are also possible \cite{dop}, $\kappa$-deformed space being one such example, where the coordinates satisfy a Lie algebra type commutation relation \cite{L1,L2,L3,L4,K1}.  Such a $\kappa$-deformed space has emerged in the attempts to construct special theory of relativity compatible with the existence of a dimensionful constant (Planck length) apart from the velocity of light in doubly special relativity \cite{D1,D2,D3}.  Apart from the studies to understand the algebraic structure and symmetries, recently, field theory models have also been investigated on such spaces \cite{F1,F2,F3,F4,F5}.

One of the notable features of field theories on noncommutative spaces with generalized symmetry is the notion of twisted statistics \cite{S1,S2,S3,S4,S5,S6}. The twisted coproduct arising from the requirement of compatibility between the algebraic structures of the noncommutative geometry and the actions of symmetry generators lead to a notion of deformed statistics. This comes about when the compatibility between the action of flip operator on multi-particle states and the twisted action of symmetry generators is demanded, leading to a twisted flip operator \cite{S3,S4}. It is this twisted flip operator that gives the definition of statistics which is invariant under the action of twisted symmetry generators. Most of these discussions reported are for the field theory models on Moyal plane, though recently, some studies have been initiated, investigating the issue of statistics for theories on $\kappa$-deformed spaces \cite{KS1,KS2,KS3,us,KS4,KS5}. Unlike the Moyal case, the twist operator in the $\kappa$-deformed space does not belong to the universal envelope of the underlying Poincar\'e algebra. Rather, the presence of the dilatation generator in the expression of the twist operator indicates that it belongs to the universal envelope of the corresponding general linear algebra \cite{lee1,lee2,lee3,jord,rim}. The twisted flip operator associated with such a twisted coproduct for the $\kappa$-deformed space was constructed in \cite{us}. In this paper we present a different proposal for a covariant flip operator for the $\kappa$-Minkowski spacetime. For this, we consider the deformed coproduct for a fully covariant realization of the $\kappa$-Minkowski algebra \cite{st1, st3}. This deformed coproduct is an element of the corresponding Hopf algebra and it is different from the twisted coproduct arising from the action of the twist operator. A covariant flip operator has to be compatible with the action of this deformed coproduct. We find an expression of such a covariant flip operator to first order in the deformation parameter, which is constructed from the generators of the Poincar\'e algebra alone. We show that the corresponding $R$-matrix also shares the same property up to first order in the deformation parameter.

In quantum field theory (QFT), the action of the twisted flip operator leads to a deformed algebra of the creation and annihilation operators. For the Moyal case, this first arose in the context of integrable models \cite{grosse} and the consequences of such a deformed oscillator algebra in QFT are well studied \cite{S1,S2,S3,S4,S5,S6}. For the $\kappa$-Minkowski case, there exists several proposals for such a deformed oscillator algebra \cite{KS3,luk1,luk2}. In this paper we construct a class of such deformed oscillator algebras corresponding to a family of realizations of the $\kappa$-Minkowski space. For particular choice of realizations, we recover the deformed oscillator algebra  obtained in \cite{KS3}, although our construction leads to a much wider class of such oscillator algebras.

This paper is organized in the following way. In Section II, we briefly review the essential details of $\kappa$-Minkowski spacetime and present a particular class of realization of the associated coordinates in terms of commuting ones and corresponding derivatives. We also present the deformed coproducts of Poincar\'e generators in terms of the functions characterizing the realization\cite{st1,st3,st4,us}. In section III, we discuss the star-product and the twist element and obtain the explicit expressions for the particular class of realization\cite{st1}. Our main results are discussed in section IV. Here, we first briefly review the twisted flip-operator for the $\kappa$-Minkowski spacetime \cite{us}. We then discuss the construction of a novel, covariant flip operator and discuss its properties.  We apply the twisted flip operator to multi-particle sector and obtain the novel, modified commutation relations between the creation and annihilation operators.  We also show how to define a new product rule between the oscillator operators so as to express their commutation relations in the familiar form. We obtain a large class of novel deformed oscillator algebras which reduce to the one discussed in \cite{KS3} for a special choice of the realization.  We finally end in section V with discussions. In the appendix, we start with the $*$-product
corresponding to the $\kappa$-spacetime that can be defined using  the commuting vectors fields and derive the twisted coproduct. Here, using this $*$-product, we first identify the twist element and using this we derive the twisted coproducts. We show that these twisted coproducts-products, for a specific realization, are exactly same as the ones we derive in section II.

\section{$\kappa$-space, Its realizations and Twisted coproducts}

In this Section, we review the results of earlier papers of some of us \cite{st1,st3,st4,us} which are used later. Similar results have been obtained in general Lie algebra type noncommutative spaces and in particular for $\kappa$ space
and quantum field theories on such spaces in \cite{L4,D2,F1,F2,F3,KS1,KS3,luk1,luk2}. Here we 
start with the generic Lie algebra type noncommutative spaces and then specialize to the case of $\kappa$-Minkowski space, for which we obtain a special class of realization of the noncommutative coordinates in terms of the coordinates and derivatives of the commuting space.

The coordinates of the generic Lie algebra type noncommutative space obey the commutation relations
\begin{equation}
[{\hat x}_\mu, {\hat x}_\nu]=iC_{\mu\nu\lambda}{\hat x}^\lambda,\quad
{\hat x}^\lambda=\eta^{\lambda \alpha} {\hat x}_\alpha
\end{equation}
with the choice $C_{\mu\nu\lambda}=a_\mu\eta_{\nu\lambda}-a_\nu\eta_{\mu\lambda}$ and $\eta_{\mu \nu} =diag(-1,1,1,\ldots 1)$ and summation over repeated indices is understood. Here $a_\mu (\mu=0,1,2,..n-1$) are real, dimensionful constants parameterizing the deformation of the Minkowski space. The $\kappa$-space is defined by the choice $a_i=0, i=1,2,...n-1;a_0=a=\frac{1}{\kappa}$. Thus we get the commutation relations between the coordinates of $\kappa$-space as
\begin{equation}
[{\hat x}_i,{\hat x}_j]=0,~~~~~[{\hat x}_0, {\hat x}_i]=ia{\hat x}_i.
\label{kcom}
\end{equation}
In terms of the Minkowski metric $\eta_{\mu\nu}={\rm diag}(-1,1,1,1......,1)$, we can define $x^{\mu} = \eta^{\mu \alpha} x_{\alpha}$ and $\partial^{\mu} = \frac{\partial }{\partial x_{\mu}} = \eta^{\mu \alpha} \partial_{\alpha}$ which satisfy the relations
\begin{equation} \label{deriv}
[ x_\mu, x_\nu] = 0, ~~~~~ [\partial_\mu, \partial_\nu] = 0, ~~~~~
[\partial^\mu, x_\nu] =  \eta_\nu^{\mu}, ~~~~~
  [\partial_\mu, x_\nu] = \eta_{\mu \nu}.
\end{equation}
For later use, we also define $p_\mu= - i\partial_\mu $ so that
$[p_\mu, x_\nu]=-i\eta_{\mu\nu}$. 

We seek realizations of the noncommutative coordinates in terms of the commuting coordinates $x_\mu$  and corresponding derivatives $\partial_\mu$ as a power series. A class of such realizations is given by 
\begin{equation}
{\hat x}_\mu=x^\alpha\Phi_{\alpha\mu}(\partial).
\end{equation}
It is easy to see that these coordinates obey $[\partial_\mu, {\hat x}_\nu]=\Phi_{\mu\nu}(\partial)$. Such a realization defines a unique mapping between the functions on noncommutative space to functions on commutative 
space. This can be seen first by defining the vacuum $|0>\equiv 1$ annihilated by $\partial$ and defining
\begin{equation}
F({\hat x}_\varphi)|0>=F_\varphi(x)\label{map}
\end{equation}
where the subscript $\varphi$ specify  the realization we work with. The functions of noncommutative coordinates are expanded as a power series in ${\hat x}_\mu$. 
Though there can be many monomials where ${\hat x}_0, {\hat  
x}_1,.....,{\hat x}_{n-1}$ appear $m_0, m_1,......,m_{n-1}$ times,  
respectively, all of them are related by the commutation relations given by Eqns. (2). Furthermore, to each  
$\varphi$-realization there exists a corresponding ordering among  
noncommutative coordinates, such that
$$:F({\hat x}_\varphi):_\varphi|0>=F(x)$$ (and vice versa).
Thus we can define left, right, totally symmetric (Weyl) ordering,  
respectively as
\begin{eqnarray}
 :e^{ik_\mu {\hat x}^\mu}:_L&\equiv& e^{-ik_0 {\hat x}_0 + ik_i{\hat 
x}_i{\varphi}_s(-ak_0)e^{-iak_0}} = e^{-ik_0 {\hat x}_0}
e^{ik_i{\hat x}_i},\nonumber\\
:e^{ik_\mu {\hat x}^\mu}:_R&\equiv &e^{-ik_0{\hat x}_0+ik_i{\hat x}_i
{\varphi}_s(-ak_0)}=e^{ik_i {\hat x}_i}e^{-ik_0{\hat x}_0},\nonumber\\
:e^{ik_\mu{\hat x}^\mu}:_S&\equiv&e^{ik_\mu{\hat x}^\mu}
\end{eqnarray}
here ${\varphi}_s(A)=\frac{A}{e^A-1}, A= ia\partial^0= -ia\partial_0$.

In this paper we work with a specific class of realization satisfying $[\partial_\mu, {\hat x}_\nu]=\Phi_{\mu\nu}(\partial)$ given by
\begin{equation}
[\partial_i,{\hat x}_j]=\delta_{ij}\varphi(A),~~~ [\partial_i, {\hat x}_0]=ia\partial_i\gamma(A),\label{pgen}
\end{equation}
\begin{equation}
[\partial_0, {\hat x}_i]=0,~~~~~~~~[\partial_0, {\hat x}_0]=\eta_{00}=-1
\end{equation}
where $A=-ia\partial_0$. An explicit solution of this realization is
\begin{eqnarray}
{\hat x}_i&=&x_i\varphi(A)\nonumber\\
{\hat x}_0&=&x_0\psi(A)+iax_i\partial_i \gamma(A). 
\label{rsation} 
\end{eqnarray}
Using the above realization in Eqn. (\ref{kcom}) we get
\begin{equation}
\frac{\varphi^\prime}{\varphi}\psi=\gamma-1
\end{equation}
where $\varphi^\prime$ is the derivative of $\varphi$ with respect to its argument $ia\partial_0$ and these functions satisfy the boundary conditions $\varphi(0)=1, \psi(0)=1$ and $\gamma(0)=\varphi^\prime(0) +1$ is finite and all are positive functions. Further demanding that the commutators of the Lorentz generators with the $\kappa$-space coordinates be linear in ${\hat x}_\mu$ as well as in the generators and have smooth commutative limit as the deformation parameter $a\rightarrow 0$ imposes further requirements on these functions and one can easily see that  there are only two class of realizations possible, viz: one where $\psi=1$ and a second one where $\psi=1+2A$. We restrict ourselves to the case $\psi=1$.

It may be noted that for $\varphi_S(A) = \frac{A}{e^A-1}$ there exists a covariant  
realization corresponding to Weyl symmetric ordering. This realization is given by
\begin{equation} \label{covreal}
\hat x_\mu = x_\mu \varphi_S(A) + i a_\mu x_\alpha \partial^\alpha \gamma_S(A), ~~~~
x_\alpha \partial^\alpha = \eta^{\alpha \beta} x_\alpha  
\partial_\beta,
\end{equation}
where $A = i a_0 \partial^0 = -i a \partial_0$. Here we choose $a_\mu = (a, 0, ..., 0)$ to be time-like, which can be chosen to be is space-like or light-like as well.

For the realization given in Eqn.(\ref{rsation}) defined by $\varphi(A)=e^{-\Lambda A/2}$,  there also exists a one-parameter family of ordering prescriptions 
\begin{equation}
 :e^{ik_\mu{\hat x}^\mu}:_{\Lambda}=e^{-i\Lambda k_0{\hat x}_0}e^{ik_i{\hat 
x}_i}e^{-i(1-\Lambda)k_0{\hat x}_0}
\end{equation}
which interpolate between right, time-symmetric and left corresponding to $\Lambda=0, \frac{1}{2},$ and $1$, respectively. Note that what we call here as totally symmetric ordering \cite{st1,st3,st4} corresponding to the realization $\varphi_S(A) = \frac{A}{e^A-1}$ is completely different from the time symmetric ordering corresponding to $\Lambda = \frac{1}{2}$ \cite{zam}.

The coproducts $\Delta_\varphi$ of the derivative operators in the realization given in Eqn. (\ref{rsation})
are
\begin{eqnarray}
\Delta_\varphi(\partial_0)&=&\partial_0\otimes I + I\otimes \partial_0~\equiv~\partial_{0}^x+\partial_{0}^y,
\label{twistcopro1}\\
\Delta_{\varphi}(\partial_i)&=&
\varphi(A\otimes I+ I\otimes A)~ \left [\frac{\partial_i}{\varphi(A)}\otimes I + e^A\otimes \frac{\partial_i}{\varphi(A)}\right].\label{twistcopro2}
\end{eqnarray}

\subsection{ $\kappa$-Poincar\'e Algebra and Casimir}

Let $M_{\mu\nu}$ denote the rotation and boost generators satisfying the undeformed  $so(n-1,1)$ algebra. We require that their commutators with the $\kappa$-space coordinates be linear functions of ${\hat x}_\mu$ and $M_{\mu\nu}$. In addition, the requirement that these commutators have a smooth commutative limit leads to
\begin{equation}
[M_{i0}, {\hat x}_0]=-{\hat x}_i +ia M_{i0}\label{gen1}
\end{equation}
\begin{equation}
[M_{i0}, {\hat x}_j]=-\delta_{ij}{\hat x}_0+ia M_{ij}.\label{gen2}
\end{equation}
 We note here that $\partial_0,\partial_i$ defined in Eqn.(\ref{pgen}) along with 
$M_{\mu\nu}$ given above generates the $\kappa$-deformed 
Poincar\'e algebra \cite{st1,st3,st4}. Note that the Lorentz algebra is undeformed and the commutator 
$[M_{\mu \nu}, \partial_\lambda] $ is deformed and depends on the realization.
We also note that the twisted coproducts of $M_{\mu\nu}$ can also be obtained from the Eqns. (\ref{gen1},\ref{gen2}) \cite{st1,st3,st4}. They are 
\begin{eqnarray}
\Delta_\varphi (M_{ij})=M_{ij}\otimes I + I\otimes M_{ij}\equiv \Delta_0(M_{ij})\label{twistcopro3}\\
\Delta_\varphi(M_{i0})=M_{i0}\otimes I + e^A\otimes M_{i0} +ia\partial_j \frac{1}{\varphi(A)}\otimes M_{ij}.\label{twistcopro4}
\end{eqnarray}

For $\psi=1$ class of realizations we are interested in, the explicit form of $M_{\mu\nu}$ are
\begin{eqnarray}
&M_{ij}=x_i\partial_j-x_j\partial_i,&\\
&M_{i0}=x_i\partial_0\varphi \frac{e^{2A}-1}{2A}-x_0\partial_i\frac{1}{\varphi}+iax_i\Delta\frac{1}{2\varphi}-iax_k\partial_k\partial_i
\frac{\gamma}{\varphi}&
\end{eqnarray}
where $\Delta=\partial_k\partial_k$. Note here that $M_{0i}$ involves $x_i\partial_i$ and so does
$\Delta_\varphi(M_{i0})$. The $\Delta (M_{i0})$ is expressed in terms of enveloping algebra of $\kappa$-deformed Poincar\'e algebra generated by $\partial_\mu, M_{\mu\nu}$.

We can also define 
\begin{equation} \label{undefromed}
\tilde M_{\mu \nu} = x_\mu \partial_\nu - x_\nu \partial_\mu = i(x_\mu  
p_\nu - x_\nu p_\mu),
\end{equation}
which generate an undeformed Poincar\'e algebra.

The generalized Klein-Gordon equation, which is invariant under the action of the undeformed Poincar\'e algebra generated by $\partial_0,\partial_i, {\tilde M}_{\mu\nu}$, is given as
\begin{equation}
(\partial_\mu\partial^\mu-m^2)\Phi(x)=0\label{undeformkg}.
\end{equation}
 Here we note that the above field equation is not invariant under the $\kappa$-deformed 
Poincar\'e transformations generated by $\partial_0,\partial_i, M_{\mu\nu}$ defined above. This can be seen easily by noticing that the derivatives do not transform like a vector under the transformations generated by $M_{\mu\nu}$. A possible way to avoid this is to introduce the (Dirac) derivatives $D_\mu$, for which there exists coordinates $X_\mu$ satisfying the conditions
\begin{equation} \label{diracdef}
[D^\mu, X_\nu] = \eta_\nu^\mu= \delta_\nu^\mu , ~~~~
[D_\mu, X_\nu] = \eta_{\mu \nu}.
\end{equation}
Then, we have, 
\begin{equation} \label{diracreal}
\hat x_\mu = X_\mu Z^{-1} + i (X_\nu a^\nu) D_\mu,~~~~
M_{\mu \nu} = X_\mu D_\nu - X_\nu D_\mu = i (X_\mu P_\nu - X_\nu  
P_\mu)  = (\hat x_\mu D_\nu - \hat x_\nu D_\mu) Z, ~~~~
P_\mu = -i D_\mu.
\end{equation}
Generators $M_{\mu \nu}$, $D_\lambda$ generate the undeformed Poincar\'e algebra.

The Dirac derivatives transform like a vector under $M_{\mu\nu}$. The
undeformed Poincar\'e algebra is defined through the relations
\begin{eqnarray}
&[M_{\mu\nu}, D_\lambda]=\eta_{\nu\lambda}D_\mu-\eta_{\mu\lambda}D_\nu,&\label{diracder1}\\
&[D_\mu,D_\nu]=0,~~~[M_{\mu\nu},\square]=0,~~~[\square, {\hat x}_\mu]=2D_\mu,&\label{diracder2}\\
&[M_{\mu\nu}, M_{\lambda\rho } ]=\eta_{\mu\rho }M_{\nu\lambda } +
\eta_{\nu\lambda}M_{\mu\rho}
  - \eta_{\nu\rho }M_{\mu\lambda } -   \eta_{\mu\lambda}M_{\nu\rho    },    \label{diracder3}&
\end{eqnarray}
which were obtained in \cite{st1,st3}. Note that $D_\mu$ and $M_{\mu\nu}$ given above generates undeformed Poincar\'e algebra. These Dirac derivatives are different from usual derivatives as can be seen easily from their action on ${\hat x}_\mu$, i,e.,
\begin{equation}
[D_\mu, \hat x_\nu] = \eta_{\mu \nu} Z^{-1} + i a_\mu D_\nu.
\label{Dderivative}
\end{equation}
where $Z^{-1}=iaD_0 +\sqrt{1+a^2D_\alpha D^\alpha}$. Using this Eqn.({\ref{Dderivative}), we get the twisted Leibniz rule for $D_\mu$ as
\begin{equation}
\Delta(D_\mu)=D_\mu\otimes Z^{-1}+I\otimes D_\mu+ia_\mu(D^\alpha Z)\otimes D_\alpha -\frac{ia_\mu}{2}\Box Z\otimes ia_\alpha  D^\alpha\label{twistcov1}.
\end{equation}
Similarly, we also get the covariant form of twisted Leibniz rule for $M_{\mu\nu}$ as
\begin{eqnarray} 
\Delta(M_{\mu\nu})&=& M_{\mu\nu}\otimes I +I\otimes M_{\mu\nu}\nonumber\\
&+&ia_\mu (D^\alpha -\frac{ia^\alpha}{2}\Box) Z\otimes M_{\alpha\nu}
-ia_\nu (D^\alpha -\frac{ia^\alpha}{2}\Box) Z\otimes M_{\alpha\mu}\label{twistcov2},
\end{eqnarray} 
where $M_{\mu\nu}$ is as given in Eqn.(\ref{diracreal}).

\subsection{Dispersion relations} 

For arbitrary realizations characterized by $\varphi$, these Dirac derivatives and $\square$ operator are
\begin{eqnarray}
D_i &=&\partial_i\frac{e^{-A}}{\varphi},\\
D_0 &=&\partial_0\frac{sinhA}{A}-ia_0\Delta\frac{e^{-A}}{2\varphi^2},\\
\square &=&\Delta\frac{e^{-A}}{\varphi^2}+2\partial_{0}^2 \frac{(1-coshA)}{{A^2}} \nonumber \label{box} \\
&=& \frac{2}{a^2}(cosh (ap_0)-1)-p_ip_i\frac{e^{-ap_0}}{\varphi^2(ap_0)}.
\end{eqnarray}
The relation between Dirac $D_\mu$ and $\partial_\mu$-derivatives  
corresponding to $\varphi_S(A)$ (Weyl symmetric ordering) is given by
\begin{equation}
D_\mu = \partial_\mu \frac{Z^{-1}}{\varphi_S(A)} + \frac{i a_\mu }{2} \Box.
\end{equation} 
and $\Box=\partial_{0}\partial^{0}\frac{e^{-A}}{(\varphi_S(A))^2}$. It is clear that the coalgebra of the undeformed Poincar\'e algebra generated by $D_\mu, M_{\mu\nu}$ is closed in enveloping algebra of Poincar\'e generators and it is a Hopf algebra.

It is also clear that the Casimir, $D_\mu D^\mu$ has vanishing commutator with $M_{\mu\nu}$ and has correct
commutative limit. The Casimir can be expressed in terms of the $\square$ operator \cite{st1,st3,st4,K1,F1} as
\begin{equation}
D_\mu D^\mu=\square(1+\frac{a^2}{4}\square).\label{casimir}
\end{equation}
Here note that the $\square$ operator is quadratic in space derivatives and thus the Casimir has quartic terms in space derivatives. 

Generalizing the notions from commutative space, it is natural to write the equation of motion for the scalar particle, i.e., generalized Klein-Gordon equation using the Casimir. Thus the generalized Klein-Gordon equation on $\kappa$-space is
\begin{equation}
(\square(1+\frac{a^2}{4}\square)-m^2)\Phi(x)=0\label{kg2}
\end{equation}
and has the correct commutative limit. But since the Casimir as well as the $\square$ operator 
have same commutative limit, the requirement of correct Klein-Gordon equation in the commutative limit does not rule out other possible generalizations in the $\kappa$-space. Thus, one can equally well start with
\begin{equation}
(\square -m^2)\Phi(x)=0\label{kg1}
\end{equation}
as the equation for scalar theory on $\kappa$-space. Other choices were also considered \cite{L1,luki1} for effective scalar Lagrangians in $\kappa$-space.

For the above choices of equations, we get the deformed dispersion relations as
\begin{eqnarray}
\frac{4}{a^2}sinh^2(\frac{ap_0}{2})-p_{i}p_{i}\frac{e^{-ap_0}}{\varphi^2(ap_0)} -m^2+\frac{a^2}{4}
\left[\frac{4}{a^2}sinh^2(\frac{ap_0}{2})-p_{i}p_{i}\frac{e^{-ap_0}}{\varphi^2(ap_0)}\right]^2=0\label{dis2}\\
\frac{4}{a^2} sinh^2(\frac{ap_0}{2}) -p_{i}p_{i}\frac{e^{-ap_0}}{\varphi^2(ap_0)} -m^2 =0,\label{dis1}
\end{eqnarray}
respectively and here $\varphi$ characterizes the realizations. Thus with $\varphi=e^{-A}, 1, \frac{A}{e^{A}-1}$ one gets left, right and Weyl-symmetric orderings, respectively. 

\section{ Star product}

The mapping between the functions on $\kappa$-space to that of commutative
space ( see Eqn. (\ref{map})) also defines a star-product, which
naturally depends on the realization $\varphi$. Thus the star-product
is defined as
\begin{equation}
F_{\varphi}({\hat x}_\varphi)G_\varphi({\hat x}_\varphi)|0>=F_\varphi *_\varphi G_\varphi.
\end{equation}
For the realizations we are interested in, i.e., the one given in Eqn.(\ref{rsation}), this implies the following $*$-product rules
\begin{eqnarray}
x_i *_\varphi f(x)= ({\hat x}_\varphi)_i f({\hat x}_\varphi)|0>=x_i\varphi(A)f(x)&\nonumber\\
x_0 *_\varphi f(x)=({\hat x}_\varphi)_0 f({\hat x}_\varphi)|0>=\left[ x_0\psi(A)+ i a x_i\partial_i \gamma(A)\right]f(x)\label{lstar}
\end{eqnarray}
and similarly
\begin{eqnarray}
f(x)*_\varphi x_i&=&x_i\varphi(A)e^Af(x)\nonumber\\
f(x)*_\varphi x_0&=&\left[x_0\psi(A)+iax_i\partial_i(\gamma(A)-1)\right]f(x).\label{rstar}
\end{eqnarray}

For any realization $\varphi$, the star-product can be expressed in terms of the twist element ${\cal F}_\varphi$ as
\begin{equation}
f*_\varphi g=m_0({\cal F}_\varphi f\otimes g)=m_\varphi (f\otimes g)\label{star1},
\end{equation}
where $f$ and $g$ are functions of the commutative coordinates and $m_0$ is the usual pointwise multiplication map in the commutative algebra of smooth functions and this can be re-expressed as
\begin{equation}
(f*_\varphi g)(x)= m_0\left( e^{x_i(\Delta_\varphi-\Delta_0)\partial_i} f(u)g(t)\right)|_{u=t=x_i},
\label{star21}
\end{equation}
where $\Delta_\varphi$ is the twisted coproduct given in Eqn. (\ref{twistcopro2}) and the undeformed coproducts is given by $\Delta_0=\partial\otimes I + I\otimes \partial.$  Comparing Eqn. (\ref{star1}) and Eqn. (\ref{star21}), we find the twist element as
\begin{equation}
 {\cal F}_\varphi = e^{x_i(\Delta_\varphi -\Delta_0)\partial_i}\label{tele}
\end{equation}
and then it is easy to find
\begin{equation}
 \Delta_\varphi={\cal F}_{\varphi}^{-1}\Delta_0{\cal F}_\varphi\label{compatibility}.
\end{equation}
 Thus we find that by applying the twist element obtained in Eqn.(\ref{tele}) to the 
undeformed coproduct of 
$\partial_0$ and  $\partial_i$, we get the twisted coproducts which are exactly same as the one obtained in Eqns.(\ref{twistcopro1},\ref{twistcopro2}). But 
${\cal F}_{\varphi}^{-1}\Delta_0({\tilde M}_{\mu\nu}){\cal F}_\varphi$
do not give the twisted coproducts of the deformed Poincar\'e algebra  obtained in Eqns.(\ref{twistcopro3},\ref{twistcopro4}), which can be easily checked using Eqn. (\ref{twistelement}) below. Also we note that the ${\tilde M}_{\mu\nu}$ along with $p_\mu$ generate undeformed Poincar\'e algebra. The corresponding coalgebra does not close in enveloping Poincar\'e algebra, but in enveloping algebra of $igl(n) \times igl(n)$. 

Using Eqn. (\ref{twistcopro2}) and Eqn. (\ref{star21}), we find that the $*$-product can be written as \cite{st4,kftmodels}
\begin{equation}
( f\star_\varphi g)(x) =e^{x_i\partial_{i}^u(\frac{\varphi(A_u+A_t)}{\varphi(A_u)}-1)+x_i\partial_{i}^t(\frac{\varphi(A_u+A_t)}
{\varphi(A_t)}e^{A_u}-1)} f(u)g(t)|_{u=t=x_i}\label{star2}.
\end{equation}
The explicit form of the corresponding twist element is now given by
\begin{equation}
 {\cal F}_\varphi=e^{N_x ln\frac{\varphi(A_x+A_y)}{\varphi(A_x)}+ N_y (A_x+ ln
\frac{\varphi(A_x+A_y)}{\varphi(A_y)})}\label{twistelement}
\end{equation}
where $N_x=x_i\frac{\partial}{\partial x_i}$ \cite{st1,st3,st4}.

Since the $*-$product depend on the ordering (or equivalently on realization) as have seen from Eqn. (\ref{lstar}) and Eqn. (\ref{rstar}), it is natural to have different twist elements depending on the ordering. Indeed, we get the twist element for left ordering as
\begin{equation}
 {\cal F}_L=e^{-N_xA_y}=e^{N\otimes A}\label{tel}
\end{equation}
and corresponding to right ordering we get
\begin{equation}
 {\cal F}_R=e^{A_xN_y}=e^{A\otimes N},\label{ter}
\end{equation}
with $A_x=-ia\partial_{0}^x$ and $N_x=x_i\partial_{i}^x$. One can combine the above two to write down an interpolating twist element
\begin{equation}
 {\cal F}_{\Lambda}=e^{-\Lambda N\otimes A+ (1-\Lambda) A\otimes N}\label{tein}
\end{equation}
which reduce to ${\cal F}_L$ and ${\cal F}_R$ when $\Lambda=1$ and $\Lambda=0$, respectively. This twist element satisfies the cocycle condition
\begin{equation}
 ({\cal F}_{\Lambda}\otimes I)(\Delta\otimes I){\cal F}_\Lambda
=(I\otimes {\cal F}_{\Lambda})(I\otimes\Delta){\cal F}_{\Lambda}.
\end{equation}
One can now get the modified momentum addition rules for
 $\kappa$-space
 from the coproducts given in Eqn. (\ref{twistcopro1}) and
 Eqn. (\ref{twistcopro2}) also. Thus going to momentum space we find
\begin{eqnarray}
\left [K_{\varphi}(p,q)\right]_\mu&=&-i\Delta_\varphi(\partial_\mu),\nonumber\\
K_\varphi(p,q) x&=& -(p_0+q_0)x_0+\varphi(-ap_0-aq_0)\left[ \frac{p_ix_i}{\varphi(-ap_0)} +\frac{e^{-ap_0}}{\varphi(-aq_0)} q_ix_i\right].
\end{eqnarray}
Similarly, we can also obtain the twist element in the momentum space, denoted by 
$ {\cal F}$ which tells how the star-product acts on the momentum space (by expressing the operators $A$ and $N$ in the momentum space in Eqns.
(\ref{tel}),(\ref{ter}) and (\ref{tein}), we get the the explicit form for ${\cal F}$, for different ordering.). Starting from 
\begin{equation}
{\cal F}f(x)g(y)\equiv {\cal F}\int d^4k d^4q e^{ikx}{\tilde f}(k) e^{iqy}{\tilde g}(q)
\end{equation}
and using the action of ${\cal F}$ on plane waves, we can easily get
\begin{equation}
{\cal F}{\tilde f}(k)\otimes{\tilde g}(q)={\cal F}(i\frac{\partial}{\partial k},k,i\frac{\partial}{\partial q},q)
{\tilde f}(k){\tilde g}(q).
\end{equation}
The above result will be of use to obtain the twisted commutation relations between the Fourier coefficients, necessary to discuss the twisted oscillators.

\section{Deformed statistics and oscillators in $\kappa$-Minkowski space}

It is known that for the QFT's defined on Moyal plane, the twisted coproduct rules affect the statistics \cite{S1,S2,S3,S4,S5,S6,us}. This is natural as the physical theory has to be invariant under the action of the underlying symmetry group of the space (or spacetime) and the definition of statistics should also be invariant under this group action. This ensures that the statistics is superselected. Such a superselection rule is implemented by demanding that the flip operator commutes with the coproduct. As the coproduct rule is now changed, we do expect a corresponding change in the definition of flip operator also. Such a twisted flip operator for the $\kappa$-deformed space was constructed in \cite{us}. In the first part of this section we briefly review that construction which requires us to consider a larger general linear algebra. Next we introduce the concept of a covariant flip operator, which preserves the algebraic structure of the $\kappa$-Minkowski space, which is a new result. We give an explicit expression of this covariant flip operator to the first order in the deformation parameter in terms of the generators of the Poincar\'e algebra alone. We also obtain an expression of the corresponding $R$-matrix to the first order. We find that upto first order in the deformation parameter, the expression for the $R$-matrix  obtained using the covariant flip operator is different from that obtained using the twisted flip operator.

Our main results are given in the second part of this section where we obtain novel twisted commutation relations between the creation and annihilation operators appearing in the mode decomposition of the scalar field satisfying generalized Klein-Gordon equation. This leads to a large class of such deformed algebras depending on the family of realizations of the $\kappa$-Minkowski space. For certain choice of realization, we explicitly obtain the deformed algebra obtained in  \cite{KS3}. Our analysis however indicates the possibility of a much wider class of deformed oscillator algebras.

\subsection{Twisted flip operators}

In this subsection, we discuss the twisted flip operators compatible with the coproducts of the deformed Poincar\'e algebra defined by the generators in Eqns.(\ref{pgen},~\ref{gen1},~\ref{gen2}) and for the undeformed Poincar\'e algebra generated by $D_\mu$ and $M_{\mu\nu}$, respectively.

In the commutative case, the flip operator is defined through its action on multi-particle states. Without loss of generality, let us consider a two-particle state $f\otimes g\in {\cal A_0}\otimes{\cal A}_0$. The action of flip operator on this (tensor product) state is given by
\begin{equation}
{\tau}_0(f\otimes g)=g\otimes f.
\end{equation}
It is easy to see that $(\tau_0)^2=I$. Symmetric and antisymmetric states of the physical Hilbert space are projected from the tensor product state as
\begin{equation}
{\frac{1}{2}} (1\pm\tau_0) (f\otimes g)= {\frac{1}{2}} (f\otimes g\pm g\otimes f),
\end{equation}
respectively. Since this definition of (anti)symmetric states should remain invariant under the action of the underlying symmetry,
its clear that the flip operator must commute with the symmetry generator. Since  $\Lambda,$ a typical element of the symmetry group acts on the tensor-product state through some representation $D$ as
\begin{equation}
\Lambda: f\otimes g= (D\otimes D)\Delta(\Lambda) f\otimes g,
\end{equation}
this requirement imply that the coproduct $\Delta(\Lambda)$ commutes with the flip operator $\tau_0$. Thus in the commutative space the flip operator $\tau_0$ is superselected so as to have vanishing commutators with all observables.
In this case of noncommutative theories, as we have seen, the coproducts get twisted and the twisted coproducts do not satisfy
\begin{equation}
[\Delta_\varphi, \tau_0]\ne 0.
\end{equation}
Thus, the meaning of (anti)symmetric states defined  using $\tau_0$ are no longer invariant. We are, thus forced to define a new twisted flip operator which commutes with the coproduct action. Since $\Delta_\varphi={\cal F}_{\varphi}^{-1}\Delta_0 {\cal F}_\varphi$, where $\Delta_0$ is the coproducts of the undeformed Poincar\'e algebra, we are immediately led to the twisted flip-operator
\begin{equation}
{\tau}_\varphi={\cal F}_{\varphi}^{-1}\tau_0 {\cal F}_\varphi\label{twistedflip}
\end{equation}
which satisfy
\begin{equation}
[\Delta_\varphi, \tau_\varphi]=0.\label{noncov}
\end{equation}
Using this twisted flip operator we can define an invariant definition of symmetric and antisymmetric states as
${\frac{1}{2}} (1\pm\tau_\varphi) (f\otimes g)$, respectively. The twisted flip operator for a generic $\varphi$ realization can be easily obtained using Eqn. (\ref{twistelement}) in Eqn. (\ref{twistedflip}) as
\begin{equation}
\tau_\varphi=e^{i(x_ip_i\otimes A-A\otimes x_ip_i)}\tau_0
\end{equation}
where $A=-ia\partial_0$. In the limit $a\rightarrow 0$, we get back the familiar commutative flip-operator, smoothly. It is interesting to note that the $\tau_\varphi$ given above is independent of $\varphi$. It may be noted that the twisted flip operator $\tau_\varphi$ is not covariant and involves operators belonging to the universal enveloping algebra of $GL(d-1,1)$. The $R$ matrix corresponding to the flip operator $\tau_\varphi$, denoted by $R_\tau$, is defined as
\begin{eqnarray}
R_\tau&=&I\otimes I +i N\wedge A\nonumber\\
&=&I\otimes I -a(x_i\partial^i)\wedge \partial_0\label{rmat}
\end{eqnarray}
and it satisfies the classical Yang-Baxter equation since $[N, A]=0$. Note that in the above $ax_i\partial^i\wedge\partial_0=a(\overleftarrow{x_i\partial^i}
\overrightarrow{\partial_0}-\overleftarrow{\partial_0}\overrightarrow{x_i\partial^i})$.

Alternately, we can define another deformed flip operator $\tau_c$  which is covariant.
This new covariant flip operator is compatible with the symmetries implemented by the covariant twisted coproducts of $D_\mu$ and $M_{\mu\nu}$ given in Eqns.(\ref{twistcov1}, \ref{twistcov2}). It is defined by the conditions
\begin{equation}
[\Delta (D_\mu), \tau_c]=0,~~  [\Delta (M_{\mu\nu}), \tau_c]=0,
\end{equation}
 where $\tau_c=R_c\tau_0$. Expanding the $R_c$-matrix in powers of the deformation parameters
$a_\mu$ as $R_c= I\otimes I+ \sum \Gamma(a, \Lambda)$, 
where $\Lambda$ stands for the generators of the $\kappa$-Poincar\'e algebra and 
using the twisted coproducts (see Eqns.(\ref{twistcov1},\ref{twistcov2})) in the 
above condition, we get the $R_c$-matrix ( to the first order in the deformation parameter) as
\begin{equation}
R_c=I\otimes I+I[M^{\mu\nu}\otimes a_\mu D_\nu -a_\mu D_\nu\otimes M^{\mu\nu}].\label{rc}
\end{equation}
The explicit form of $M_{\mu\nu}$ appearing above is given in Eqn.(\ref{diracreal}).
We note here that the above $R$ matrix, up to first order in the parameter involves only the generators
of the $\kappa$-Poincar\'e algebra, namely $M_{\mu\nu}$ and $D_\mu$. This has to be contrasted with the one in Eqn.(\ref{rmat}) for the non-covariant, twisted flip operator which involves (space) dilation operator which is not in the k-Poincar\'e algebra($R$-Matrix, as an expansion in inverse powers of $\kappa$ was studied in \cite{yz}
for the case $\kappa$ deformed spaces.) It may also be noted that the $R_\tau$ and $R_c$ matrices would in general lead to different physics. The calculation of the covariant $R$-matrix to all orders in the deformation parameter is presently under investigation. This covariant $R$-matrix to first order in the deformation parameter given in Eqn.(\ref{rc}) is a new result.

\subsection{Twisted Oscillator algebra}

In this section, we derive a novel class of twisted products between the creation and annihilation operators appearing in the mode expansion of the scalar field theory in $\kappa$-space. 

Having defined the twisted flip operator $\tau_\varphi$, we are now in a position to define (anti) symmetric states of a theory defined in the $\kappa$-Minkowski space. We start by defining the deformed bosonic state as
\begin{equation}
f\star_\varphi g = m_\varphi(f\otimes g)=m_\varphi\tau_\varphi(f\otimes g).
\end{equation}
Using the definitions of $m_\varphi$ and $\tau_\varphi$( see Eqn. (\ref{star1}) and Eqn. (\ref{twistedflip})) in the
above, we get
\begin{equation}
f\otimes g=\tau_\varphi (f\otimes g)\label{twistedboson}
\end{equation}
or equivalently we can write
\begin{equation}
{\cal F}_\varphi(f\otimes g)={\tilde{\cal F}}_\varphi(f\otimes g)
\end{equation}
where we have used the mirror twist operator ${\tilde{\cal F}}_\varphi=\tau_0{\cal F}_\varphi {\tau_0}$. Now defining the twisted tensor product $f\otimes_\varphi g$ as ${\cal F}_\varphi(f\otimes g)$, from the above, we get
\begin{equation}
f\otimes_\varphi g=\tau_0(f\otimes_\varphi g).
\end{equation}

For the product of two bosonic fields $\phi(x)$ and $\phi(y)$ under interchange, now we pick up an additional factor compared to the commutative case. This can be calculated using Eqn. (\ref{twistedboson}) and one gets
\begin{equation}
\phi(x)\otimes\phi(y)-e^{-(A\otimes N-N\otimes A)}\phi(y)\otimes\phi(x)=0.
\end{equation}
Expressing $\phi$ in the above equation using Fourier transforms and using the twisted flip operator in momentum space, we are led to the deformed commutation relations between the annihilation operators as
\begin{equation}
{\tilde\phi}(k){\tilde\phi}(p)=e^{-ia\left[ k_0(\partial_{p_i} p_i)-p_0(\partial_{k_i}k_i)\right]}
{\tilde\phi}(p){\tilde\phi}(k)\label{twistcom}
\end{equation}

The $\Phi(x)$ appearing in the generalized Klein-Gordon equation (\ref{undeformkg}) can be expressed as
\begin{equation}
\Phi(x)=\int d^4p \delta(p_{0}^2-\omega^2){\bar A}(p) e^{-ip\cdot x}
\end{equation}
where $\omega=\sqrt{p_{i}^2+m^2}$.
Using the mode decomposition
\begin{equation}
\Phi(x)=\int \frac{d^3 p}{\sqrt{p_{i}^2+m^2}} \left[A(\omega, {\vec p}) e^{-ip\cdot x}+A^\dagger(\omega, {\vec p})e^{ip\cdot x}\right],
\end{equation}
 from Eqn.(\ref{twistcom}) we get
\begin{eqnarray}
A^\dagger(p_0, {\vec p})A(q_0, {\vec q})-e^{-a(q_0\partial_{p_i}p_i+\partial_{q_i} q_i p_0)} 
A(q_0,{\vec q})A^\dagger(p_0,{\vec p})=-\delta^3(p-q),\label{modcom1}\\
A^\dagger(p_0,{\vec p})A^\dagger(q_0,{\vec q})-e^{-a(-q_0\partial_{p_i}p_i+\partial_{q_i}q_i p_0)} A^\dagger(q_0,{\vec q})A^\dagger(p_0,{\vec p})=0,\label{modcom2}\\
A(p_0,{\vec p})A(q_0,{\vec q})-e^{-a(q_0\partial_{p_i}p_i-\partial_{q_i} q_i p_0)} A(q_0,{\vec q})A(p_0,{\vec p})=0.\label{modcom30}
\end{eqnarray}

For the choice $\varphi=e^{-\frac{A}{2}}=e^{-\frac{ia\partial_0}{2}}$, the generalized 
Klein-Gordon equation (\ref{kg1}) is
\begin{equation}
\left[\partial_{i}^2 +\frac{4}{a^2} sinh^2(\frac{ia\partial_0}{2})-m^2\right]\Phi=0.\label{dkg}
\end{equation}
We can decompose this field in positive and negative frequency modes and thus
\begin{equation}
\Phi(x)=\int \frac{d^4p}{2\Omega_{k}(p)}\left[
A(\omega_{k},{\vec p}) e^{-ip\cdot x} +A^\dagger(\omega_{k},{\vec p}) e^{ip\cdot x}\right]\label{pmmodes}
\end{equation}
where $A^\dagger(\pm\omega_{k},{\vec p})=A^\dagger(\mp\omega_{k},{\vec p})$.
In the above, we have used
\begin{eqnarray}
p_{0}^\pm=\pm \omega_{k}(p)=\pm \frac{2}{a}sinh^{-1}(\frac{a}{2}\sqrt{p_{i}^2+m^2}),\label{P0}\\
\Omega_k(p)=\frac{1}{a} sinh(a\omega_k(p)).\label{omg}
\end{eqnarray}

Using this in Eqn.(\ref{twistcom}), for the field satisfying the deformed generalized Klein-Gordon equation
given above ( Eqn.(\ref{dkg})), we obtain various twisted commutation relations between creation and annihilation operators. They are
\begin{eqnarray}
A^\dagger(p_0, {\vec p})A(q_0, {\vec q})-e^{-a(q_0\partial_{p_i}p_i+\partial_{q_i} q_i p_0)} 
A(q_0,{\vec q})A^\dagger(p_0,{\vec p})=-\delta^3(p-q),\label{modcom11}\\
A^\dagger(p_0,{\vec p})A^\dagger(q_0,{\vec q})-e^{-a(-q_0\partial_{p_i}p_i+\partial_{q_i}q_i p_0)} A^\dagger(q_0,{\vec q})A^\dagger(p_0,{\vec p})=0,\label{modcom21}\\
A(p_0,{\vec p})A(q_0,{\vec q})-e^{-a(q_0\partial_{p_i}p_i-\partial_{q_i} q_i p_0)} A(q_0,{\vec q})A(p_0,{\vec p})=0\label{modcom31}
\end{eqnarray}
Note that $p_0$ and $q_0$ are as given in Eqn. (\ref{P0}).
From this, one can easily derive the following relations
\begin{eqnarray}
&A^\dagger(p_0, e^{-\frac{aq_0}{2}}{\vec q})A^\dagger(q_0, e^{\frac{ap_0}{2}}
{\vec q})-{\cal F}(q,p)A^\dagger(q_0, e^{-\frac{ap_0}{2}}{\vec q})A^\dagger(p_0, e^{\frac{aq_0}{2}}{\vec q})=0,&\\
&A(p_0, e^{\frac{aq_0}{2}}{\vec p}) A(q_0,e^{-\frac{ap_0}{2}}{\vec q})
-{\cal F}(-q,-p)A(q_0,e^{\frac{ap_0}{2}}{\vec q})A(p_0,e^{-\frac{aq_0}{2}}{\vec p})=0,&\\
&A^\dagger(p_0, e^{\frac{aq_0}{2}}{\vec p})A(q_0, e^{\frac{ap_0}{2}}{\vec q})-{\cal F}(-q,p)
A(q_0,e^{-\frac{ap_0}{2}}{\vec q})A^\dagger(p_0, e^{-\frac{aq_0}{2}}{\vec p})=-
\delta^3(p-q).&
\end{eqnarray}
where ${\cal F}(q,p)=e^{3a(q_0-p_0)}.$
These relations were obtained in \cite{KS3,luk2} using different approach. Using these relations, a new product (the $\circ$-product) between the creation and annihilation operators is defined as follows
\begin{eqnarray}
A(p)\circ A(q)=e^{-\frac{3a}{2}(p_0-q_0)}
A(p_0, e^{\frac{aq_0}{2}}{\vec p})A(q_0, e^{-\frac{ap_0}{2}}{\vec q})\label{twpr1}\\
A^\dagger(p)\circ A^\dagger(q)=e^{\frac{3a}{2}(p_0-q_0)}
A^\dagger(p_0, e^{-\frac{aq_0}{2}}{\vec p})A(q_0, e^{\frac{ap_0}{2}}{\vec q})\\
A^\dagger(p)\circ A(q)=e^{\frac{3a}{2}(p_0+q_0)}
A^\dagger(p_0, e^{\frac{aq_0}{2}}{\vec p})\circ A(q_0,e^{\frac{ap_0}{2}}{\vec q})\\
A(p)\circ A^\dagger(q)=e^{-\frac{3a}{2}(p_0+q_0)}
A(p_0, e^{-\frac{aq_0}{2}}{\vec p})\circ A^\dagger(q_0, e^{-\frac{ap_0}{2}}{\vec q})\label{twpr2}.
\end{eqnarray}
Using this new product rule, we can re-express Eqns.(\ref{modcom11},\ref{modcom21},\ref{modcom31}) as
\begin{equation}
[A(p_0,{\vec p}), A(q_0,{\vec q})]_\circ=0,~~~[A^\dagger(p_0,{\vec p}), A^\dagger(q_0,{\vec q})]_\circ=0,\label{twcom1}
\end{equation}
\begin{equation}
[A(p_0,{\vec p}),A^\dagger(q_0,{\vec q})]_\circ=\delta^3({\vec p}-{\vec q}).\label{twcom2}
\end{equation}
Thus, with this modified product rule, the algebra of creation and annihilation operators can be recast in the same form as the corresponding commuting operators.

We note here that the creation and annihilation operators satisfying the specific deformed products given in
Eqns.(\ref{twpr1}-\ref{twpr2})( and hence the commutation relations in Eqns.(\ref{twcom1}, \ref{twcom2})) are 
the ones appearing in the mode decomposition of the scalar field (see Eqn.(\ref{pmmodes})) satisfying the generalized Klein-Gordon Eqn. (\ref{kg1}) with a particular choice $\varphi(A)=e^{-\frac{A}{2}}$( see Eqn.({\ref{dkg})).
Thus it is clear that even for the scalar field obeying the field Eqn.(\ref{kg1}), more general (i.e., for other choices of/arbitrary
$\varphi(A)$) dispersion relations than those given in Eqns.(\ref{P0},\ref{omg}) are possible. This will lead to more general twisted products than those given in Eqns.(\ref{modcom1}-\ref{twpr2}), leading to generalized commutation relations in place of those in Eqns.(\ref{twcom1},\ref{twcom2}). Thus the twisted products (see Eqns.(\ref{twpr1}-\ref{twpr2})) obtained in \cite{KS3,luk2} are only a {\it particular case} of more general products between $A^\dagger$ and $A$ that are possible.

The creation and annihilation operators satisfying above given deformed commutation 
relations are the ones appearing in the mode decomposition of the scalar field satisfying the generalized Klein-Gordon Eqn.(\ref{kg1}). This generalized Klein-Gordon equation is invariant under the  $\kappa$-Poincar\'e algebra defined in Eqns.(\ref{diracder1},\ref{diracder2}), in addition to the usual $so(n-1,1)$ commutation relations between $M_{\mu\nu}$. This should be contrasted with the approach taken in \cite{KS3,luk2} where the generalized Klein-Gordon equation was invariant under the action of a {\it deformed} $\kappa$-Poincar\'e algebra. Irrespective of this, we have obtained the deformed commutation relations between  $A^\dagger$ and $A$ given in \cite{KS3,luk2}, as a special case.

\section{Conclusion}

In this paper, we have studied the construction of scalar theory on $\kappa$-Minkowski spacetime. It is known that this noncommutativity
of the coordinates leads to twisted coproducts for the generators of the $\kappa$-Poincar\'e algebra. These twisted coproducts are necessary for the implementation of the symmetry algebra on multiparticle states. We have summarized briefly, explicit form of the twisted coproducts for a class of realization of  deformed $\kappa$-spacetime coordinates in terms of commuting coordinates and derivatives. Here, the momenta do not transform like a vector unlike in the case of undeformed $\kappa$-Poincar\'e algebra. This results in the non-invariance of the naive generalization of generalized Klein-Gordon equation (see Eqn.(\ref{undeformkg})) under the action of the deformed $\kappa$-Poincar\'e algebra. We then introduced Dirac derivatives which transform as vector under the deformed $\kappa$-Poincar\'e algebra. After obtaining the coproducts of the generators of this deformed Poincar\'e algebra and the Casimir, generalized Klein-Gordon equations which are invariant under this algebra
are introduced. The requirement of invariance alone does not lead to a unique generalized Klein-Gordon equation. These generalized Klein-Gordon equations do have  higher derivative terms with respect to time while one of them has quartic space derivatives (see Eqn. (\ref{kg2})) while the second has quadratic space derivatives( see Eqn. (\ref{kg1})).  We have then discussed the $*$-product naturally induced by the realization of $\kappa$-spacetime coordinates in terms of the commuting ones and derivatives. From this star product, one can read-off the twist element and we showed that it can be expressed in terms of space dilation operator and time derivative. It is clear that the twisted coproducts obtained using this twist element are different from those of deformed $\kappa$-Poincar\'e algebra as well as those of the undeformed $\kappa$-Poincar\'e algebra defined using Dirac derivatives.

Our main results are discussed in section $IV$. Here we have derived the flip operators compatible with the algebraic structure of the system.  First we have obtained the twisted flip operator which is compatible with the twisted coproducts of the deformed $\kappa$-Poincar\'e algebra and then we derive the covariant flip operator which is compatible with the coproducts of the undeformed $\kappa$-Poincar\'e algebra defined using Dirac derivatives. In both cases, we have obtained the R-matrices corresponding to the deformed flip operators (up to first order in the deformation parameter). It is shown that in the first case, the twisted flip operator contains elements that do not belong to the set of the generators of the symmetry algebra. In contrast, the covariant flip operator up to the first order in the deformation parameter involves only the generators of the symmetry algebra. The calculation of the covariant R-matrix to all orders in the deformation parameter is being investigated now. Whether the two different $R$-matrices we obtained for the $\kappa$-Minkoswksi space are equivalent or not is under investigation.

We have then studied implications of the twisted flip operator on the statistics of the scalar field quanta, satisfying generalized Klein-Gordon equation defined in the $\kappa$-spacetime. We have shown that the algebra of creation and annihilation operators is deformed and we obtain this deformed algebra explicitly. We have also shown that this deformed algebra reproduces a known result for a specific choice of the realization of the $\kappa$-spacetime coordinates \cite{KS3}. Our analysis however leads to a much wider and novel class of deformed oscillator algebras.

Finally, in the appendix, we have discussed the $*$-product for the $\kappa$ spacetime defined using vector fields and obtain the twisted coproducts of the symmetry algebra induced by this $*$-product. We show that this twisted coproduct is same as that of the undeformed $\kappa$-Poincar\'e algebra generated by $\partial_\mu,$ and ${\tilde M}_{\mu\nu}$.

\vskip 1cm
\noindent{\bf ACKNOWLEDGEMENTS}\\

TRG, KSG and EH would like to thank J. Lukierski for useful discussions. This work was supported by the Ministry of Science and Technology of the Republic of Croatia under contract No. 098-0000000-2865. This work was done within the framework of the Indo-Croatian Joint Programme of Cooperation in Science and Technology sponsored by the Department of Science and Technology, India (DST/INT/CROATIA/P-4/05), and the Ministry of Science, Education and Sports, Republic of Croatia.

\appendix
\section {The Star Product and Coproducts from Vector fields} 

Here we discuss the $*$-product and the twist element for the$\kappa$-space-time using the commuting vector field. Using this twist element, we then derive the twisted coproducts and show that they are same as the ones we obtained in Eqns.(\ref{twistcopro1},\ref{twistcopro2},\ref{twistcopro3},\ref{twistcopro4}), with a specific  choice $\varphi(a)$.

In section 3, we have obtained the $*$-product induced on the commutative space by the mapping of function from deformed k-space-time. The explicit form of this $*$-product is given in Eqn.(\ref{star2}). $*$-product was first introduced in quantum mechanics as a way to handle the ordering problems encountered in the passage from classical phase space to quantum one( or in obtaining a map between functions on
classical phase space to the corresponding quantum mechanical operators). The approach, later lead to the development of deformation quantization. In the studies of non-commutative field theories on Moyal plane, it was realized that the product of functions on Moyal plane can be replaced with $*$-product of function on commutative space. Later, it was shown that the compatibility requirement of $*$-product and the action of symmetry generators introduces the twisted coproducts into the discussions of symmetries\cite{chaichan}.

The $*$-product in the Moyal plane is constructed using the commuting generators of the symmetry algebra( namely Poincar\'e algebra). Later $*$-products were constructed for quantum spaces like $M(so_q(3))$, $M(so_q(1,3))$ etc using commutative vector fields ( which are not the generators of the symmetry algebra of the underlying space)\cite{sykora}. For a pair of vector fields ${\bf X}$ and ${\bf Y}$, this $*$-product is given a series. An asymmetric $*$-product is defined by
\begin{equation}
f*g=\sum_{n=0}^\infty \frac{h^n}{n!} ({\bf X}^n f)({\bf Y}^n g)
\end{equation}
and a symmetric one is defined as
\begin{equation}
f*g=\sum_{n=0}^\infty\frac{h^n}{2^n n!}\sum_{i=0}^n (-1)^n~nC_i
({\bf X}^{n-i}{\bf Y}^i f)({\bf X}^{i}{\bf Y}^{n-i} g)\label{symmstar}
\end{equation}

Using this approach, one can define a $*$-product between the coordinates leading to the commutation relations defining the$\kappa$-space-time given in Eqn.(\ref{kcom}). This product is defined in terms of the commuting vector fields $x_i\partial_i$ and $\partial_0$ as
\begin{equation}
f*g=f e^{-i\frac{a}{2}(x_i\partial_i\otimes \partial_0-\partial_0\otimes x_i\partial_i)} g.\label{kstar}.
\end{equation}
Here we have summed the series in Eqn.(\ref{symmstar}) to get the phase factor in the above equation.
It is easy to verify that the above star product gives the Eqn.(\ref{kcom}) for commuting coordinates $x_0, x_i$.

As in the Moyal case, here too, we can define the twisted coproduct in terms of the twist element ${\cal F}$. Noting $m_*(fg)=f*g=m({\cal F}fg)$ and using Eqn.(\ref{kstar}), we find ${\cal F}=e^{-i\frac{a}{2}(x_i\partial_i\otimes \partial_0-\partial_0\otimes x_i\partial_i)}$. Using $\Delta_t(g)={\cal F}^{-1}\Delta(g){\cal F}$, we find the twisted coproducts of   the generators of the undeformed Poincar\'e algebra $\partial_0, \partial_i, {\tilde M}_{0i},$ and ${\tilde M}_{ij}$. We find
\begin{eqnarray}
\Delta_t(\partial_0)&=&\Delta(\partial_0);~\Delta_t({\tilde M}_{ij})=\Delta({\tilde M}_{ij})\\
\Delta_t(\partial_i)&=&\partial_i\otimes e^{-\frac{ia\partial_0}{2}}
+e^{\frac{ia\partial_0}{2}}\otimes \partial_i\\
\Delta_t({\tilde M}_{i0})&=&-i[x_i\partial_0\otimes e^{\frac{ia\partial_0}{2}} +
e^{-\frac{ia\partial_0}{2}}\otimes x_i\partial_0]-i[
x_0\partial_i\otimes e^{-\frac{ia\partial_0}{2}}+e^{\frac{ia\partial_0}{2}}\otimes x_0\partial_i]\nonumber\\
&+&\frac{a}{2}[\partial_i\otimes x_j\partial_j e^{-\frac{ia\partial_0}{2}}
-x_j\partial_j e^{\frac{ia\partial_0}{2}}\otimes\partial_i].
\end{eqnarray}
Exactly the same twisted coproducts result from 
${\cal F}_{\varphi}^{-1}\Delta_0(\partial_\mu){\cal F}_\varphi,~$
${\cal F}_{\varphi}^{-1}\Delta_0(M_{\mu\nu}){\cal F}_\varphi$ where $\Delta_0$ is the coproducts of the undeformed Poincar\'e generators with the choice $\varphi(a)=e^{-\frac{ia\partial_0}{2}}$ ( for the derivative operators these are same as the ones given in Eqns.(\ref{twistcopro1},\ref{twistcopro2})). We note that the $*$ product defined using the commutating vector fields in Eqn.({\ref{kstar}) is same as the one we obtained in Eqn.(\ref{star2}) and the twist element is exactly same as that in Eqn.(\ref{twistelement}) with the choice $\varphi(a)=e^{-\frac{ia\partial_0}{2}}$. Thus it should not be surprising that these two different approaches leads to same twisted coproducts. But it is clear the $*$ product defined in Eqn.({\ref{star2}) is more general as it reduces to Eqn.(\ref{kstar}) only for the choice $\varphi(a)=e^{-\frac{ia\partial_0}{2}}$. We also note that the twisted coproducts derived above are given in terms of operators that are not in $\kappa$-Poincar\'e algebra.

\end{document}